\documentclass[preprintnumbers,amsmath,amssymb,floatfix,9pt,prd,twocolumn,showpacs,superscriptaddress,nofootinbib]{revtex4}
\usepackage{graphicx}
\usepackage{epsfig}
\usepackage{bm}
\usepackage{amsfonts}
\def\be {\begin{equation}}
\def\ee {\end{equation}}
\def\ba {\begin{eqnarray}}
\def\ea {\end{eqnarray}}
\begin{document}

\title{Thermodynamics of an Evolving Lorentzian Wormhole with Entropy Corrections}

\author{Tanwi Bandyopadhyay}\email{tanwib@gmail.com}
\affiliation{Department of Mathematics,~Shri~Shikshayatan
College, 11, Lord~Sinha Road,~Kolkata-71, India}

\author{Ujjal Debnath} \email{ujjaldebnath@yahoo.com,ujjal@iucaa.ernet.in}
\affiliation{Department
of Mathematics,~Bengal Engineering and Science
University,~Shibpur,~Howrah-711103, India.}

\author{Mubasher Jamil} \email{mjamil@camp.nust.edu.pk}
\affiliation{Center for Advanced Mathematics and Physics (CAMP),
National University of Sciences and Technology (NUST), H-12,
Islamabad, Pakistan}
\affiliation{Eurasian International Center
for Theoretical Physics, Eurasian National University, Astana
010008, Kazakhstan}

\author{Faiz-ur-Rahman}
\affiliation{Center for Advanced Mathematics and Physics (CAMP),
National University of Sciences and Technology (NUST), H-12,
Islamabad, Pakistan}

\author{Ratbay Myrzakulov}
\email{rmyrzakulov@gmail.com;
rmyrzakulov@csufresno.edu}\affiliation{Eurasian International Center
for Theoretical Physics, Eurasian National University, Astana
010008, Kazakhstan}

\begin{abstract}
 {\bf Abstract:} In this work, we study the
generalized second law of thermodynamics (GSL) at the apparent
horizon of an evolving Lorentzian wormhole. We obtain the
expressions of thermal variables at the apparent horizon. Choosing
the two well-known entropy functions i.e. power-law and
logarithmic, we obtain the expressions of GSL. We have analyzed
the GSL using a power-law form of scale factor $a(t)=a_0t^n$ and
the special form of shape function $b(r)=\frac{b_{0}}{r}$. It is
shown that GSL is valid in the evolving wormhole spacetime for
both choices of entropies if the power-law exponent is small, but
for large values of $n$, the GSL is satisfied at initial stage and
after certain stage of the evolution of the wormhole, it
violates.

\end{abstract}

\maketitle

\newpage

\section{\normalsize\bf{Introduction}}

Different theories of quantum gravity particularly string theory
and loop quantum gravity predict the thermal nature of black hole
horizons \cite{r1,r2,s00}. These theories confirm the
semi-classical results of Bekenstein and Hawking about the
evaporation of black holes as a result of emission of charged
particles \cite{h1,h2}. In the early seventies, it was found that
laws of thermodynamics are applicable to black holes horizons by
defining new quantities like entropy ($S\sim A$) and surface
gravity ($\kappa\sim T$) \cite{h3}. Hence the laws of
thermodynamics for black holes arise from the geometry of
spacetime. More recently in the last few years, several new
connections between geometry/gravity and thermodynamics have been
found. Jacobson found a link between the Einstein equation and the
thermal relation $\delta Q=TdS$, where left hand side represents
the energy flux \cite{j2} (see excellent reviews on connections
between gravity and thermodynamics \cite{rev1,rev2,rev3}). Later
it was proved that the thermodynamics of black holes is also valid
for cosmological horizons \cite{davies} and the Friedmann equation
can be written as a first law of thermodynamics
\cite{cai,akbar1,akbar2}. While more recently it is proposed by
Verlinde that gravity has an entropic origin and Newton's law of
gravitation arises naturally and unavoidably in a theory in which
space is emergent through a holographic scenario
\cite{verlinde}.

Wormholes are tunnels in spacetime geometry that connect two or more
regions of the same spacetime or two different spacetimes
\cite{Morris}. Wormholes are classified in two categories: Euclidean
wormholes that arise in Euclidean quantum gravity and the Lorentzian
wormholes which are static spherically symmetric solutions of
Einstein's general relativistic field equations \cite{Visser1}.
Interest in Lorentzian wormholes arose in order to make them useful
for time travel for human beings, hence termed Time Machines in
popular literature \cite{thorne}. In order to support such exotic
wormhole geometries, the matter required to stabilize them has to be
exotic i.e. violating the energy conditions (null, weak and strong),
however it has been shown that averaged null energy condition is
satisfied in wormhole geometries \cite{visser,w1,w2}. Also the
violation of weak energy condition (WEC) can be avoided for small
intervals of time \cite{viss}. More recently traversable wormhole and time machine solutions of the field equations of an alternative of gravity with non-minimally curvature-matter coupling have been obtained \cite{fer}. Wormholes have been constructed in some alternative theories of gravity including $f(R)$, teleparallel, Kaluza-Klein, loop quantum gravity, Brans-Dicke gravity and Lovelock gravities, to name a few \cite{extra} and is found that modified gravity can minimize the amount of anisotropy and exotic matter required to support the existence of wormhole's throat.

Kar \& Sahdev had shown that if the
wormhole spacetime is taken to be evolving (non-static) than the
matter threading the wormhole geometry need not to be WEC violating
\cite{kar}. In their model, the wormhole initially inflates and
later evolves like a FRW spacetime thereby implying our Universe to
be an evolving wormhole. Later Anchordoqui et al \cite{Anchordoqui}
extended the previous work \cite{kar} by choosing the non-zero
redshift function and obtained evolving wormhole solutions with
non-exotic matter. Arellano \& Lobo \cite{lobo} found contracting
wormhole solutions within non-linear electrodynamics. Cataldo et al
 discussed the construction of evolving wormholes
using phantom energy \cite{Cataldo,cat} and more recently using
the cosmological constant \cite{cat2}. Farooq et al \cite{Farooq}
showed that the field equations of a (2+1)-dimensional evolving
wormhole can be recast into first law of thermodynamics. Debnath et al \cite{deb} investigated the non-static Lorentzian wormhole model in presence of anisotropic pressure and showed that the Einstein's field equations and unified first law are equivalent for the dynamical wormhole model.  Inspired
by the above works, we are interested to study the thermodynamics
of an evolving wormhole at its apparent horizon using general
entropy function and two special entropy functions, to be
discussed later.

The paper is organized as follows: In section II, we write down
the governing dynamical equations of an evolving wormhole. The
previous work of Rahaman et al \cite{Ak} has some computational
errors for wormhole thermodynamics in presence of isotropic
pressure. In this work, we correct these assigning with
anisotropic pressure in the field equations. In section-III, we
study the wormhole thermodynamics by deriving the expressions of
temperature, entropy and surface gravity at the apparent horizon
of the wormhole. Further we obtain the general expression of the
GSL at the apparent horizon valid for any type of entropy function
chosen. In section IV, we study the expression of GSL by choosing
two well-known forms of entropy functions i.e. log corrected and
the power-law corrected in two forthcoming subsections. We analyze
the behavior of GSL by choosing a power-law form of scale factor.
We briefly discuss our results in section V.

\section{\normalsize\bf{Basic equations }}

A simple generalization of Morris-Thorne (MT) wormhole to the
time-dependent background is given by the evolving Lorentzian
wormhole \cite{Cataldo}
\begin{equation}\label{1}
ds^{2}=-e^{2\Phi(t,r)}dt^{2}+a^{2}(t)\Big[\frac{dr^{2}}
{1-\frac{b(r)}{r}}+r^{2}d\Omega_{2}^{2}\Big],
\end{equation}
where $d\Omega_{2}^{2}=d\theta^{2}+\sin^{2}\theta d\phi^{2}$ and
$a(t)$ is the scale factor of the universe. The functions $b(r)$
and $\Phi(t,r)$ are called the shape function and redshift
function respectively. Earlier the evolving wormholes have been studied with time-independent redshift function $\Phi(r)$ \cite{cat}. If $a(t)\rightarrow$ constant and
$\Phi(t,r)\rightarrow\Phi(r)$, then the static MT wormhole can be
obtained whereas for $b(r)\rightarrow kr^{3}$ and
$\Phi(t,r)\rightarrow$ constant, Eq (\ref{1}) reduces to standard FRW
metric.

For anisotropic pressure the components of the energy-momentum
tensor are \cite{cat,cat2} assumed as
\begin{equation}\label{2}
T^{t}_{t}=-\rho(t,r),T^{r}_{r}=-p_{r}(t,r),T^{\theta}_{\theta}=T^{\phi}_{\phi}=-p_{t}(t,r),
\end{equation}
where, the traditionally homogeneous and isotropic exotic matter
source has been generalized to an inhomogeneous and anisotropic
fluid, but still with a diagonal energy-momentum tensor (as is
usually considered in phantom cosmologies). Here $\rho(t,r)$,
$p_{r}(t,r)$ and $p_{t}(t,r)$ are the energy density, radial and
tangential pressures respectively. The Einstein's field equations
are then given by
\begin{eqnarray}
3e^{-2\Phi}H^{2}+\frac{b'}{a^{2}r^{2}}&=&8\pi G\rho,\label{3}\\
-e^{-2\Phi}(2\dot{H}+3H^{2})+2e^{-2\Phi}\dot{\Phi}H-\frac{b}{a^{2}r^{3}}&=&8\pi
Gp_{r},\label{4}\\
-e^{-2\Phi}(2\dot{H}+3H^{2})+2e^{-2\Phi}\dot{\Phi}H+\frac{b-rb'}{2a^{2}r^{3}}&=&8\pi
Gp_{t},\label{5}\\
2\Phi'H&=&0,\label{6}
\end{eqnarray}
where $H=\frac{\dot{a}}{a}$ is the Hubble parameter. Here dot and dash
stand for the differentiation with respect to $t$ and $r$
respectively. From eqn (\ref{6}) we have $\Phi'=0~\rightarrow
\Phi(t,r)=\Phi(t)$. So without any loss of generality, we can set
$\Phi=0$ by re-scaling the time coordinate \cite{Cataldo}. Then the field eqns
change to
\begin{eqnarray}
3H^{2}+\frac{b'}{a^{2}r^{2}}&=&8\pi G\rho,\label{7}\\
2\dot{H}+3H^{2}+\frac{b}{a^{2}r^{3}}&=&-8\pi Gp_{r}, \label{8}\\
2\dot{H}+3H^{2}-\frac{b-rb'}{2a^{2}r^{3}}&=&-8\pi Gp_{t},\label{9}
\end{eqnarray}
From eqns (\ref{8}) and (\ref{9}), we have
\begin{equation}\label{10}
\frac{3b-rb'}{2a^{2}r^{3}}=-8\pi G(p_{t}-p_{r}).
\end{equation}
Also, from the conservation of energy equation, we get
\begin{equation}\label{11}
\dot{\rho}+H(3\rho+p_{r}+2p_{t})=0,
\end{equation}
\begin{equation}\label{12}
2(p_{t}-p_{r})=rp_{r}'.
\end{equation}
For isotropic pressure, $p_{r}=p_{t}$ and hence from eqn (\ref{12}),
$p_{r}'=0$ i.e, $p_{r}$ becomes a function of time only. Also eqn
(\ref{10}) shows $b(r)=kr^{3}$ and thus the metric (\ref{1}) becomes FRW metric.
To avoid this structure, we must require matter with anisotropic
pressure which automatically leads to inhomogeneity.

\section{\normalsize\bf{Wormhole Thermodynamics}}

Some basic laws of wormhole dynamics have been derived in
\cite{Hayward}, which suggest a genuine connection with
thermodynamics. Ever since, the thermal properties of wormholes
have been studied in literature \cite{Hong,Bokhari,Martin}. For
studying the generalized second law of thermodynamics for an
evolving wormhole, let us consider the metric (\ref{1}) as in the
following form
\begin{equation}\label{13}
ds^{2}=h_{ij}dx^{i}dx^{j}+\tilde{r}^{2}d\Omega_{2}^{2},~~~~i,j=0,1
\end{equation}
where
$h_{ij}=\Big(-1,a^{2}\Big(1-\frac{b(r)}{r}\Big)^{-1}\Big)$
and the physical radius $\tilde{r}=a(t)r$.\\

The dynamical apparent horizon (AH) $\tilde{r}_{A}$ is given by
\begin{equation}\label{14}
[h^{ij}\partial_{i}\tilde{r}\partial_{j}\tilde{r}]_{\tilde{r}=\tilde{r}_{A}}=0,
\end{equation}
i.e.,
\begin{equation}\label{15}
H^{2}\tilde{r}_{A}^{2}=1-\frac{ab\Big(\frac{\tilde{r}_{A}}{a}\Big)}{\tilde{r}_{A}}.
\end{equation}
Differentiating (15), we have
\begin{equation}\label{16}
\dot{\tilde{r}}_{A}=\frac{H\tilde{r}_{A}\Big[\tilde{r}_{A}b'\Big(\frac{\tilde{r}_{A}}{a}\Big)
-ab\Big(\frac{\tilde{r}_{A}}{a}\Big)-2\dot{H}\tilde{r}_{A}^{3}\Big]}{\Big[\tilde{r}_{A}b'\Big(\frac{\tilde{r}_{A}}{a}\Big)
-ab\Big(\frac{\tilde{r}_{A}}{a}\Big)+2H^{2}\tilde{r}_{A}^{3}\Big]}.
\end{equation}
The associated surface gravity is defined as
\begin{equation}\label{17}
\kappa=\frac{1}{2\sqrt{-h}}~\partial_{i}(\sqrt{-h}~h^{ij}\partial_{j}\tilde{r}),
\end{equation}
i.e.,
\begin{equation}\label{18}
\kappa=-\frac{\tilde{r}_{A}}{2}(\dot{H}+2H^{2})+\frac{1}{4\tilde{r}_{A}^{2}}
\Big[ab\Big(\frac{\tilde{r}_{A}}{a}\Big)-\tilde{r}_{A}b'\Big(\frac{\tilde{r}
_{A}}{a}\Big)\Big].
\end{equation}
Hence the associated temperature on the AH is
\begin{eqnarray}\label{19}
T_{A}=\frac{\kappa}{2\pi}&=&-\frac{\tilde{r}_{A}}{4\pi}(\dot{H}+2H^{2})\nonumber\\&&+\frac{1}{8\pi\tilde{r}_{A}^{2}}
\Big[ab\Big(\frac{\tilde{r}_{A}}{a}\Big)-\tilde{r}_{A}b'\Big(
\frac{\tilde{r}_{A}}{a}\Big)\Big].
\end{eqnarray}
Now, we assume a functional form of the entropy on the AH as in
the following form \cite{Chak}
\begin{equation}\label{20}
S_{A}=\frac{f(A)}{4G},
\end{equation}
where $A=4\pi\tilde{r}_{A}^{2}$ and  $f(A)$ is the entropy function. Then
\begin{equation}\label{21}
\frac{dS_{A}}{dt}=\frac{2\pi}{G}f'(A)\tilde{r}_{A}\dot{\tilde{r}}_{A}.
\end{equation}
Then from (\ref{19}) and (\ref{21}), we may write
\begin{eqnarray}\label{22}
T_{A}\frac{dS_{A}}{dt}&=&\frac{2\pi}{G}f'(A)\tilde{r}_{A}\dot{\tilde{r}}_{A}
\Big[-\frac{1}{4\pi}\tilde{r}_{A}(\dot{H}+2H^{2})\nonumber\\&&+\frac{1}{8\pi\tilde{r}_{A}^{2}}
\Big\{ab\Big(\frac{\tilde{r}_{A}}{a}\Big)-\tilde{r}_{A}b'\Big(\frac{
\tilde{r}_{A}}{a}\Big)\Big\}\Big].
\end{eqnarray}
On the other hand, for anisotropic pressure the Gibb's equation
can be defined as
\begin{equation}\label{23}
T_{A}dS_{I}=\frac{1}{3}(p_{r}+2p_{t})dV+d(\rho V),
\end{equation}
where $S_{I}$ is the entropy within the AH and
$V=\frac{4}{3}\pi\tilde{r}^{3}$. So, it can be shown that
\begin{equation}\label{24}
T_{A}\frac{dS_{I}}{dt}=\frac{4\pi\tilde{r}_{A}^{2}}{3}\Big(3\rho+p_{r}+2p_{t}
+\frac{\tilde{r}_{A}\rho'}{a}\Big)\Big(\dot{\tilde{r}}_{A}-H\tilde{r}_{A}\Big).
\end{equation}
Thus from (\ref{22}) and (\ref{24}), the variation of total entropy on the AH
can be expressed as
\begin{eqnarray}\label{25}
T_{A}\Big(\frac{dS_{A}}{dt}+\frac{dS_{I}}{dt}\Big)&=&\frac{2\pi}{G}f'(A)\tilde{r}_{A}\dot{\tilde{r}}_{A}
\Big[-\frac{1}{4\pi}\tilde{r}_{A}(\dot{H}+2H^{2})\nonumber\\&&+\frac{1}{8\pi\tilde{r}_{A}^{2}}
\Big\{ab\Big(\frac{\tilde{r}_{A}}{a}\Big)-\tilde{r}_{A}b'\Big(\frac{\tilde{r}_{A}
}{a}\Big)\Big\}\Big]\nonumber\\&&+\frac{4\pi\tilde{r}_{A}^{2}}{3}\Big(3\rho+p_{r}+2p_{t}
+\frac{\tilde{r}_{A}\rho'}{a}\Big)\nonumber\\&&\times\Big(\dot{\tilde{r}}_{A}-H\tilde{r}_{A}\Big).
\end{eqnarray}
If $\Big(\frac{dS_{A}}{dt}+\frac{dS_{I}}{dt}\Big)>0$ i.e., if
the expression of the r.h.s of above equation is non-negative,
then we say that GSL is valid. For this purpose we need to know
the entropy function $f(A)$ and the shape function $b(r)$.\\

\section{Two Special Cases}

Here we choose a particular form of the shape function as
$b(r)=\frac{b_{0}}{r}$, $b_{0}=$~constant, in order to study the
GSLT for two different types of wormhole models by choosing two
specific forms of the horizon entropy $S_{A}$ i.e., logarithmic
and power law correction entropies.

\subsection{\normalsize\bf{Logarithmic Correction}}

It is a well-known fact that in Einstein's gravity, the entropy of
the horizon is proportional to the area, but when the gravity theory
is modified by adding extra curvature terms in the action, it also
modifies the entropy-area relation. In the context of loop-quantum
gravity, this relation can be expanded into an infinite series as
\cite{Banerjee}
\begin{equation}\label{26}
S_{A}=S_{0}+\tilde{\alpha}~\text{ln}S_{0}-\sum_{i=1}^{\infty}\frac{\tilde{\alpha_{i}}}
{S_{0}^{i}}.
\end{equation}
Here $S_{0}$ is the classical entropy and the higher order terms
are quantum corrections. Here $\alpha_{i}$'s are finite constants,
but their values are highly debatable. For example, some take the
value of $\tilde{\alpha}$ to be negative \cite{Kaul, Ghosh}, some
positive \cite{Hod}, whereas some have taken it to be zero
\cite{Medved}. This correction has been used in literature for
variety of purposes: to study the dark energy models in numerous
gravities \cite{j,f0,f2}, to unify inflation and dark energy
models \cite{j0}, study of GSL with log corrections
\cite{j1,j4}.\\

From Eq (\ref{26}), we continue our study by considering the expression of
the horizon entropy with the logarithmic correction only \cite{Wei},
i.e, for our study
\begin{equation}\label{27}
S_{A}=\frac{A}{4G}+\pi\beta~\text{ln}\Big(\frac{A}{4G}\Big),
\end{equation}
so that the entropy function can be written as
\begin{equation}\label{28}
f(A)=A+4\pi G\beta~\text{ln}\Big(\frac{A}{4G}\Big).
\end{equation}

With this choice of entropy, eqn (\ref{25}) takes the form
\begin{eqnarray}\label{29}
T_{A}\Big(\frac{dS_{A}}{dt}+\frac{dS_{I}}{dt}\Big)&=&-\frac{H\tilde{r}_{A}}{2}
\Big(\frac{1}{G}+\frac{\beta}{\tilde{r}_{A}^{2}}\Big)\Big(\frac{a^{2}b_{0}
+\dot{H}\tilde{r}_{A}^{4}}{H^{2}\tilde{r}_{A}^{4}-a^{2}b_{0}}\Big)\nonumber\\&&\times\Big[\frac{a^{2}b_{0}}{\tilde{r}_{A}^{2}}
-\tilde{r}_{A}^{2}(\dot{H}+2H^{2})\Big]\\&&
-\frac{H\tilde{r}_{A}^{7}}{3G}\Big(\frac{a^{2}b_{0}}{\tilde{r}_{A}^{4}}
-3\dot{H}\Big)\Big(\frac{H^{2}+\dot{H}}{H^{2}\tilde{r}_{A}^{4}-a^{2}b_{0}}\Big),\nonumber
\end{eqnarray}
where we have used the field equations and eqn (\ref{16}) for the
expressions of $\rho$, $p_{r}$, $p_{t}$ and
$\dot{\tilde{r}}_{A}$.\\

For a special choice of the scale factor in power law form which
is assumed as $a(t)=a_{0}t^{n}$. Now, we plot the evolution of the
total entropy on the AH against time in figure 1, where the
constant $n$ has been taken to be equal to $2$ (solid line) and
$5$ (dashed line), we see that the GSL is valid for small $n$ but
as $n$ increases, GSL is not obeyed in the later epoch for
logarithmic corrected entropy.

\begin{figure}
\includegraphics[scale=0.8]{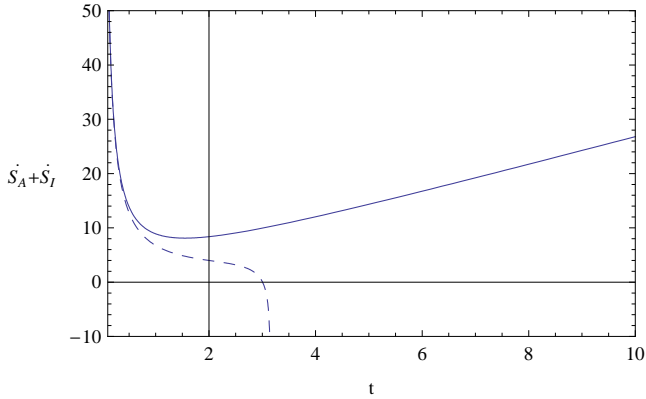}
\caption{ This figure represents the rate of change of
total entropy for logarithmic correction of the horizon entropy
function against time for $n=2$ (solid line) and $n=5$ (dashed line).}
\end{figure}

\subsection{\normalsize\bf{Power Law Correction}}

Another significant correction term to the horizon entropy
expression appeared while dealing with the entanglement of quantum
fields inside and outside of the horizon where the wave function of
the field is chosen to be a superposition of ground state and
excited state \cite{Das}. So the associated power law corrected
entropy expression becomes \cite{Radicella,jamil}
\begin{equation}\label{30}
S_{A}=\frac{A}{4G}\Big(1-K_{\alpha}A^{\frac{1-\alpha}{2}}\Big),
\end{equation}
so that in this case we have
\begin{equation}\label{31}
f(A)=A\Big(1-K_{\alpha}A^{\frac{1-\alpha}{2}}\Big),
\end{equation}
where $\alpha$ is a dimensionless constant and $K_{\alpha}$ is given
by
\begin{equation}\label{32}
K_{\alpha}=\frac{\alpha(4\pi)^{\frac{\alpha}{2}-1}}{(4-\alpha)r_{c}^{2-\alpha}}.
\end{equation}
Here $r_{c}$ is called the cross-over scale. The entanglement
entropy of the ground state obeys the usual area law whereas the
excited state contributes to the correction. Thus the correction
term is more significant for higher excitations. It is important
to note that the correction term falls off rapidly as $A$
increases and hence in the semi classical limit (large $A$), the
area law can be recovered. This correction has been recently used
extensively in dark energy literature: to study the holographic
and new-agegraphic dark energy in various gravitational theories
\cite{jamil1,jamil2,jamil3,karami,k,s,k1} and the study of GSL in
FRW cosmology with power-law entropy correction
\cite{debnath,k3}.

For this choice of the horizon entropy, eqn (25) becomes
\begin{eqnarray}\label{33}
T_{A}\Big(\frac{dS_{A}}{dt}+\frac{dS_{I}}{dt}\Big)&=&-\frac{H\tilde{r}_{A}^{(2-\alpha)}}{4G}
\Big[2\tilde{r}_{A}^{(\alpha-1)}+(\alpha-3)K_{\alpha}(4\pi)^{\frac{1-\alpha}{2}}\Big]\nonumber\\&&
\Big(\frac{a^{2}b_{0}+\dot{H}\tilde{r}_{A}^{4}}{H^{2}\tilde{r}_{A}^{4}-a^{2}b_{0}}\Big)
\Big[\frac{a^{2}b_{0}}{\tilde{r}_{A}^{2}}-\tilde{r}_{A}^{2}(\dot{H}+2H^{2})\Big]\nonumber\\&&
-\frac{H\tilde{r}_{A}^{7}}{3G}\Big(\frac{a^{2}b_{0}}{\tilde{r}_{A}^{4}}
-3\dot{H}\Big)\Big(\frac{H^{2}+\dot{H}}{H^{2}\tilde{r}_{A}^{4}-a^{2}b_{0}}\Big).
\end{eqnarray}
Here also the field equations together with eqn (\ref{16}) have been
used to write the expressions of $\rho$, $p_{r}$, $p_{t}$ and
$\dot{\tilde{r}}_{A}$. For a special choice of the scale factor in
power law form which is assumed as $a(t)=a_{0}t^{n}$. Now, we plot
the evolution of the total entropy on the AH against time in
figure 2, where the constant $n$ has been taken to be equal to $2$
(solid line) and $5$ (dashed line), we see that the GSL is valid
for small $n$ but as $n$ increases, GSL is not obeyed in the later
epoch for power law corrected entropy. Hence the behavior is
analogous for both power-law and Log-corrected entropies.

\begin{figure}
\includegraphics[scale=0.8]{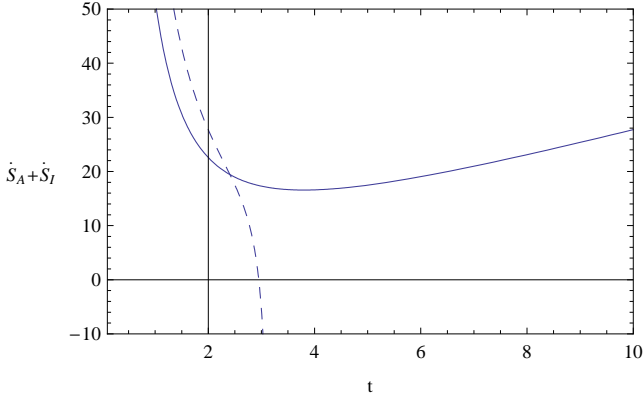}
\caption{ This figure represents the rate of change of
total entropy for power law corrected form of the horizon entropy
function against time for $n=2$ (solid line) and $n=5$ (dashed line).}
\end{figure}

\section{Exact solutions}

Let us find some exact solutions of the system (\ref{7})-(\ref{9}).

i) Let $a=a_0t^n$, then $H=nt^{-1}$.  Substituting these expressions into (\ref{7})-(\ref{9}) we get (below we assume $8\pi G=1$)
\begin{eqnarray}
\rho&=&\frac{3n^2}{t^{-2}}+\frac{b'}{a^{2}_0t^{2n}r^{2}},\label{34} \\
p_r&=&\frac{n(2-3n)}{t^2}-\frac{b}{a^{2}_0t^{2n}r^{3}},\label{35} \\
 p_T&=&\frac{n(2-3n)}{t^2}+\frac{b-rb'}{2a^{2}_0t^{2n}r^{3}}.\label{36}
\end{eqnarray}
Now let us introduce the EoS parameters as
\be
\omega_r(r,t)=\frac{p_r}{\rho},\quad
\omega_T(r,t)=\frac{p_T}{\rho}.\label{37}
 \ee
 For the solutions (\ref{34})-(\ref{36}) we obtain
 \begin{eqnarray}
\omega_r&=&\frac{n(2-3n)a^{2}_0r^{3}t^{2n-2}-b}{r[3n^2a^{2}_0r^{2}t^{2n-2}+b^{'}]},\label{38} \\
\omega_T&=&\frac{2n(2-3n)a^{2}_0r^{3}t^{2n-2}+(b-rb^{'})}{2r[3n^2a^{2}_0r^{2}t^{2n-2}+b^{'}]}. \label{39}
\end{eqnarray}
In these formulas we have one arbitrary function $b(r)$ and parameters $n, \quad a_0$. Let us we assume that in the $r$-direction we have  an accelerated expansion so that we can put $\omega_r=-1$. Then from (\ref{38}) we determine the unknown function $b(r)$  as
\be
b(r)=r[C-na^{2}_0r^{2}t^{2n-2}],\label{40}
 \ee
 where $C=constant$. To get rid of the $t$ dependence of $b$ we put $n=1$. Then finally from (\ref{40}) we get
 \be
b(r)=r[C-a^{2}_0r^{2}],\label{41}
 \ee
 So for the EoS parameters we get
  \begin{eqnarray}
\omega_r&=&-1,\label{42} \\
\omega_T&=&0.  \label{43}
\end{eqnarray}

ii) Our next example is by taking $\omega_r=const=\omega_{r0}$. Then for the density of energy, pressures and EoS parameters we get the same  expressions as in the previous case. We get $b(r)$ from (\ref{38})
\be
b(r)=Cr^{-\frac{1}{\omega_{r0}}}+\frac{2n-3(1+\omega_{r0})n^2}{1+3\omega_{r0}}a^{2}_0r^{3}t^{2n-2},\label{44}
 \ee
 where $C=constant$. To get rid of the $t$ dependence of $b$ we put $n=1$. Then finally we get
 \be
b(r)=Cr^{-\frac{1}{\omega_{r0}}}-a^{2}_0r^{3},\label{45}
 \ee
 Following \cite{w2}, we can check the stability conditions for this evolving wormhole. These conditions include certain restrictions on the form of shape function including (1) $b'(r_0)<1$, at the wormhole's throat (2) $b(r)<r$ for $r>r_0$ (flare-out condition) and (3) $b(r)/r\rightarrow0$, as $|r|\rightarrow\infty$ (asymptotic flatness). Clearly the last condition for (\ref{45}) is not satisfied, however the first two condition impose the following inequality $$C<r^{1+\frac{1}{w_{r0}}}(1+a_0^2r^2).$$

 In our case the formulas (\ref{38})-(\ref{39}) become
  \begin{eqnarray}
\omega_r&=&-\frac{a^{2}_0r^{3}+b}{r[3a^{2}_0r^{2}+b^{'}]},\label{46} \\
\omega_T&=&\frac{b-rb^{'}-2a^{2}_0r^{3}}{2r[3a^{2}_0r^{2}+b^{'}]}.  \label{47}
\end{eqnarray}
So from these formulas and (\ref{45}) finally we get
  \begin{eqnarray}
\omega_r&=&\omega_{r0},\label{48} \\
\omega_T&=&-\frac{1+\omega_{r0}}{2r}. \label{49}
\end{eqnarray}
Consider particular cases. 1) Let $\omega_{r0}=1/3$ that is raditation. Then
 \begin{eqnarray}
\omega_r&=&1/3,\label{50} \\
\omega_T&=&-\frac{4}{6r}.  \label{51}
\end{eqnarray}
The solution suggests a radiation state parameter in the $r$ direction while a phantom energy state parameter $r<\frac{2}{3}$.

2) Let $\omega_{r0}<-1$ that is phantom matter. Then
 \begin{eqnarray}
\omega_r&<&-1,\label{52} \\
\omega_T&>&0. \label{53}
\end{eqnarray}
3) Let $\omega_{r0}>1$ that is ekpyrotic matter. Then
 \begin{eqnarray}
\omega_r&>&1,\label{54} \\
\omega_T&<&-\frac{1}{r}.  \label{55}
\end{eqnarray}
It is interesting to note that in this case we have the ekpyrotic matter in $r$-direction and phantom in $T$-direction if $r<1$. So that $r=1$ is a cruical value.

iii) Now let us consider the case
\be
\omega_{r0}=\frac{2-3n}{3n}.\label{56}
 \ee
In this case $b(r)$ takes the form
\be
b(r)=Cr^{\frac{3n}{3n-2}},\label{57}
 \ee
Then the expressions for the parameters of EoS (\ref{38})-(\ref{39}) become
 \begin{eqnarray}
\omega_r&=&\frac{2-3n}{3n},\label{58} \\
\omega_T&=&-\frac{1}{3n}\left[1+\frac{3n(1-n)(2-3n)a^{2}_0r^{2}t^{2n-2}}{Cr^{\frac{2}{3n-2}}+n(3n-2)a^{2}_0r^{2}t^{2n-2}}\right].  \label{59}
\end{eqnarray}
From (\ref{58}), it is easy to see that a cosmological constant state parameter cannot be obtained. In other words, an evolving wormhole cannot be constructed and supported from vacuum energy.

\begin{figure}
\includegraphics[scale=0.4]{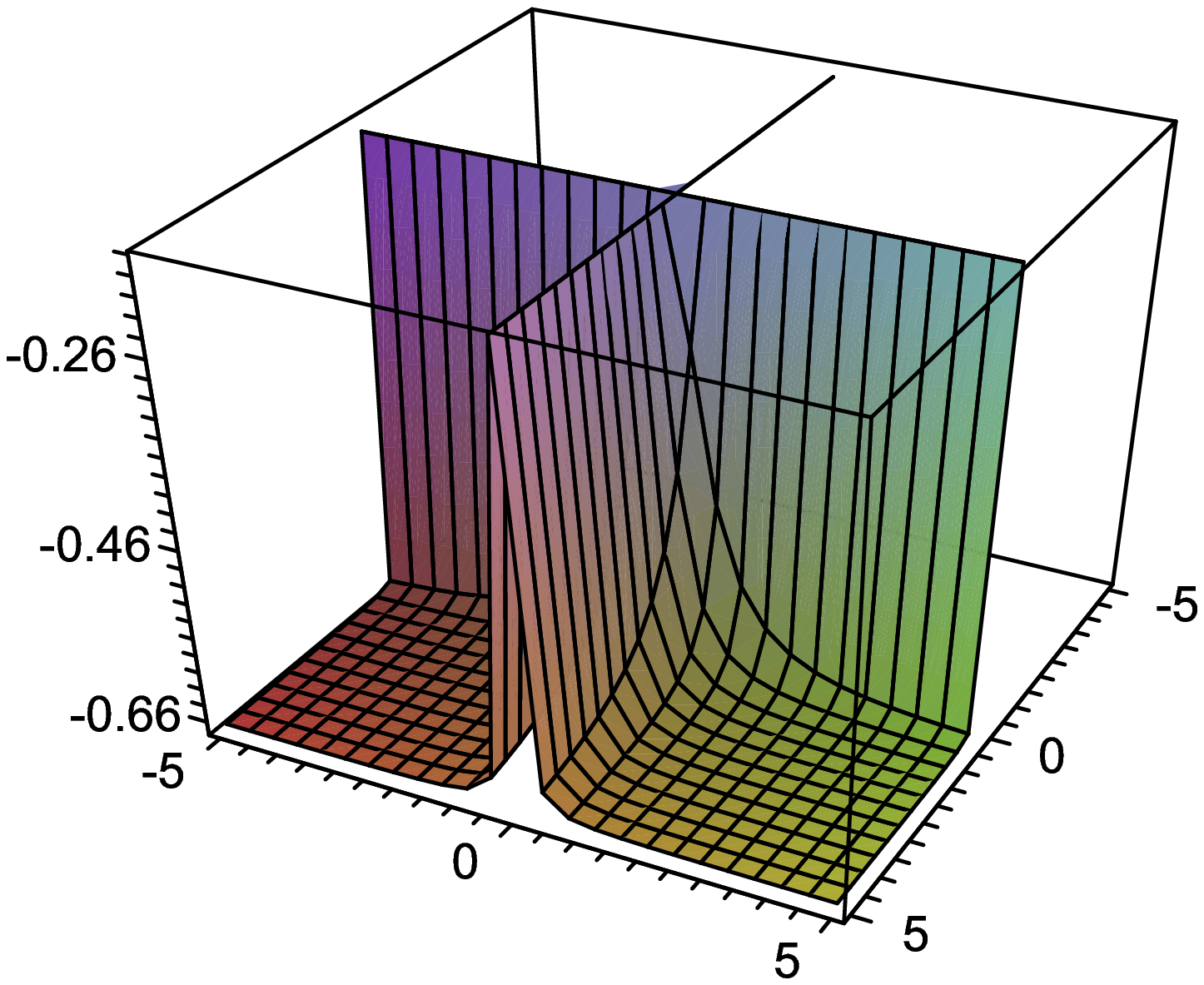}
\caption{ This figure shows the evolution of $w_T$ (Eq.\ref{59})} against $r$ and $t$, both having same range on the horizontal axes. The model paramters are fixed at $a_0=1$, $C=1$, $n=2$.
\end{figure}

\section{\normalsize\bf{Conclusions}}

In this work, we have studied the basic equations of evolving
Lorentzian wormhole by assuming the anisotropic pressure. The
generalized second law of thermodynamics (GSL) at the apparent
horizon of an evolving Lorentzian wormhole has been analyzed in
general way when we have considered the horizon entropy
proportional to a function of the horizon area. We have obtained
the expressions of thermal variables at the apparent horizon.
Choosing the two well-known entropy functions i.e. power-law and
logarithmic, we have obtained the expressions of the variation of
the total entropy using Gibb's equation. We have analyzed the GSL
using a simple well established power-law form of scale factor
$a(t)=a_0t^n$ and the special form of shape function
$b(r)=\frac{b_{0}}{r}$. It is shown that GSL is valid in the
evolving wormhole spacetime for both choices of entropies if the
power-law exponent $n$ is small, but for large values of $n$, the
GSL is satisfied at initial stage and after certain stage of the
evolution of the wormhole, it violates.

\subsection*{Acknowledgements}

One of the authors (TB) wants to thank UGC, Govt. of India for
providing with a research project No. F.PSW-063/10-11 (ERO). The
authors (TB, UD) are thankful to IUCAA, Pune, India for warm
hospitality where part of the work was carried out.

\end{document}